\documentclass[utf8]{frontiersSCNS} 

\usepackage{url,lineno,microtype,subcaption}
\usepackage[onehalfspacing]{setspace}
\usepackage[finalnew]{trackchanges}

\def\keyFont{\fontsize{8}{11}\helveticabold }
\def\firstAuthorLast{Oliveira and Zesta} 
\def\Authors{Denny M. Oliveira\,$^{1,2,*}$, Eftyhia Zesta$^2$, and Sergio Vidal-Luengo$^3$}

\def\Address{$^{1}$Goddard Planetary Heliophysics Institute, University of Maryland, 
				   Baltimore County, Baltimore, MD, United States
			 $^{2}$Geospace Physics Laboratory, NASA Goddard Space Flight Center, Greenbelt, MD, United States
			 $^3$Laboratory for Atmospheric and Space Physics, University of Colorado, Boulder, CO, United States
			 }

\def\corrAuthor{Denny M. Oliveira}
\def\corrEmail{denny@umbc.edu}

\def\thxn{$\theta_{x_n}$}
\def\dbdt{d$B$/d$t$}
\def\dbxdt{d$B_x$/d$t$}

\usepackage[breaklinks, colorlinks = true,
            linkcolor = SkyBlue,
            urlcolor = SkyBlue,
            citecolor = SkyBlue,
            anchorcolor = white,
            breaklinks]{hyperref}

\def\man{M\"ants\"al\"a}

\begin{document}
\onecolumn
\firstpage{1}

\title[Shock-induced GICs at \man]{First direct observations of interplanetary shock impact angle effects on actual geomagnetically induced currents: The case of the Finnish natural gas pipeline system} 

\author[\firstAuthorLast ]{\Authors} 
\address{} 
\correspondance{} 

\extraAuth{}


\maketitle

\begin{abstract}

	The impact of interplanetary (IP) shocks on the Earth's magnetosphere can greatly disturb the geomagnetic field and electric currents in the magnetosphere-ionosphere system. At high latitudes, the current systems most affected by the shocks are the auroral electrojet currents. These currents then generate ground geomagnetically induced currents (GICs) that couple with and are highly detrimental to ground artificial conductors including power transmission lines, oil/gas pipelines, railways, and submarine cables. Recent research has shown that the shock impact angle, the angle the shock normal vector performs with the Sun-Earth line, plays a major role in controlling the subsequent geomagnetic activity. More specifically, due to more symmetric magnetospheric compressions, nearly frontal shocks are usually more geoeffective than highly inclined shocks. In this study, we utilize a subset (332 events) of a shock list with more than 600 events to investigate, for the first time, shock impact angle effects on the subsequent GICs right after shock impact (compression effects) and several minutes after shock impact (substorm-like effects). We use GIC recordings from the Finnish natural gas pipeline performed near the \man{} compression station in southern Finland. We find that GIC peaks ($>$ 5 A) occurring after shock impacts are mostly caused by nearly frontal shocks and occur in the post-noon/dusk magnetic local time sector. These GIC peaks are presumably triggered by partial ring current intensifications in the dusk sector. On the other hand, more intense GIC peaks ($>$ 20 A) generally occur several minutes after shock impacts and are located around the magnetic midnight terminator. These GIC peaks are most likely caused by intense energetic particle injections from the magnetotail which frequently occur during substorms. The results of this work are relevant to studies aiming at predicting GICs following solar wind driving under different levels of asymmetric solar wind forcing.

	\tiny
 	\keyFont{ \section{Interplanetary shocks, shock geometry, geomagnetic activity, geospace response, ionospheric response, geomagnetically induced currents}} 
\end{abstract}

\section{Introduction}

	Interplanetary (IP) shocks correspond to a kind of perturbation frequently observed in the solar wind at many locations in the heliosphere \citep{Smith1985a,Szabo2001,Aryan2014,Echer2019a,Perez-Alanis2023}. IP shocks are formed when the \change{relative speed between the shock structure and the local solar wind velocities}{relative speed between the Rankine-Hugoniot-determined shock velocity and the upstream solar wind velocity} is greater than the environment magnetosonic speed \citep{Priest1981,Kennel1985,Parks2004,Piel2010}. This results from sharp enhancements of solar wind plasma properties (velocity, number density, temperature) and interplanetary magnetic field, known as IMF, characterizing the formation of fast forward shocks \citep{Tsurutani2011a,Oliveira2017a}. IP shocks are expected to occur during all phases of the solar cycle, but they are much more common during solar maxima \citep{Kilpua2015a,Oliveira2015a,Rudd2019}. Strengths of IP shocks are usually represented by magnetosonic Mach numbers, the ratio between the shock/solar wind relative speed and the local magnetosonic speed \citep{Tsurutani1985b,Lugaz2016,Oliveira2017a}. IP shocks are usually driven by coronal mass ejections \citep[CMEs,][]{Tsurutani1988,Veenadhari2012} and corotating interaction regions \citep[CIRs,][]{Smith1976,Fisk1980}. \par

	The impact of IP shocks on the magnetosphere often causes geomagnetic disturbances observed in the geospace, ionosphere, and on the ground. Such responses are characterized by magnetic field disturbances at geosynchronous orbit \citep{Kokubun1983,Nagano1984,Wing1997}, field-aligned currents in the magnetosphere-ionosphere system \citep{Moretto2000,Araki2009,Belakhovsky2017,Liu2023}, sudden impulses observed in ground magnetometer data \citep{Echer2005a,Shinbori2009,Wang2010d}, and magnetospheric substorms triggered by explosive energy release by the Earth's magnetotail \citep{Kokubun1977,Zhou2001,Milan2017}. More important for this work, IP shocks trigger ground \dbdt{} variations that can be observed at high latitudes \citep{Pulkkinen2017,Ngwira2018c}, mid latitudes \citep{Marshall2011,Fiori2014}, low/equatorial latitudes \citep{Carter2015,Nilam2023}, \add{and everywhere} \citep{Tsurutani2014a}. Such field variations connect to ground conductors through geoelectric fields according to Faraday's law ($\vec{\nabla}\times\vec{E} = - \partial\vec{B}/\partial t$) which in turn generate geomagnetically induced currents (GICs) \citep{Boteler1998,Viljanen1998,Tsurutani2014a}. Therefore, ground \dbdt{} variations \add{and auroral electroject variations (in the east-west and other directions)} are recognized as the space weather drivers of GICs \citep{Dimmock2019,Tsurutani2023,Wawrzaszek2023}. However, a model of the Earth's conductivity must be used in order to characterize GICs triggered by enhanced geoelectric fields \citep{Viljanen2006a,Pulkkinen2007a,Bedrosian2015}. GIC effects can be detrimental to artificial conductors found in ground power transmission lines \citep{Erinmez2002,Trivedi2007,Pulkkinen2017,Piccinelli2018}, \add{old telegraph wires} \citep{Barlow1849,Arcimis1903,Hayakawa2020a}, oil and gas pipelines \citep{Campbell1980,Martin1993,Gummow2002}, railways \citep{Kasinskii2007,Love2019b,Patterson2023}, and submarine cables \citep{Charkraborty2022,Boteler2024}. \par 
 
 	Geomagnetic activity triggered by IP shocks is significantly controlled by shock impact angles, which correspond to the angle the normal vector performs with the Sun-Earth line. More specifically, the more frontal the shock impact, the higher the subsequent geomagnetic activity. In general, \add{for Earth-bound shocks observed at L1} CME-driven shocks have their shock normals aligned with the Sun-Earth line due to radial CME propagation \citep{Klein1982,Gulisano2010,Salman2020}, whereas CIR-driven shocks are more inclined due to the twisted nature of the Parker spiral \citep{Pizzo1991,Jian2006a,Rout2017}. \add{\protect{Thus, possible shock normal deviations caused by interplanetary medium variations, such as magnetic field and plasma density \citep{Temmer2023}, are neglegible.}} Many numerical and experimental studies have shown that shocks with small inclinations tend to trigger more intense field-aligned currents \citep{Guo2005,Selvakumaran2017,Shi2019b}; cause sudden impulse events with shorter rise times \citep{Takeuchi2002b,Wang2006a,Rudd2019}, and can determine whether substorms are triggered or not \citep{Oliveira2014b,Oliveira2015a,Oliveira2021b}. In addition, \cite{Oliveira2018b} showed that nearly frontal and high-speed shocks drive more intense ground \dbdt{} variations at all latitudes right after shock impacts. \cite{Oliveira2021b} showed in a comparative study that a nearly frontal shock triggered a substorm much more intense than a substorm triggered by a highly inclined shock, even though both shocks had similar strengths. They attributed these observations to the fact that the magnetosphere was more rapidly and symmetrically compressed in the nearly frontal shock case, while the compression was slower and asymmetric in the highly inclined case. As a result, ground \dbdt{} variations were more intense, occurred earlier, and covered larger geographic areas in North America and Greenland as indicated by a large array of ground magnetometers. These results were confirmed by the superposed epoch analysis study reported by \cite{Oliveira2024a} with similar data. In general, most works agree that more frontal shocks, in comparison to inclined shocks, tend to compress the magnetosphere more symetrically enhancing current systems in the magnetosphere and ionosphere more effectively, which in turn leads to higher geomagnetic activity, as reviewed by \cite{Oliveira2018a} and more recently by \cite{Oliveira2023b} \par

 	\cite{Viljanen2010} reported on a statistical study of GICs in southern Finland covering approximately one solar cycle. The authors catalogued the highest GIC amplitudes in the period of 1999 to 2010. \cite{Viljanen2010} concluded that the GIC peaks occurred mostly during intense magnetic storms near solar maximum. \cite{Tsurutani2021} used the same GIC data set to investigate the solar wind and magnetospheric conditions associated with GIC peaks larger than 30 A in a more \change{elastic}{extensive} period (1999 to 2019). In that work, it was concluded that such extremely high GIC peaks mostly occurred during magnetospheric super substorms, which take place when ground magnetometers show intense activity of the westward auroral electrojet with lower envelope indices $< -$2,500 nT \citep{Tsurutani2015,Hajra2018a,Zong2021}. Although these works advanced our understanding of GIC enhancements at high latitudes during magnetic storm times, they did not provide a link between magnetospheric compressions caused and substorms triggered by IP shocks with different inclinations. As we will show in this paper, shock-induced GICs can also pose serious threats to artificial conductors in long-, mid-, and short-term regimes. The results presented in this work also have important implications to GIC forecasting, since IP shock impact angles can be forecasted in a time window of up to 2 hours before shock impacts on the Earth's magnetosphere \citep{Paulson2012,Oliveira2018b,Smith2020a}. \par

 	Although some works have shown that shock impact angles significantly affect ground \dbdt{} variations, direct shock impact angle effects on actual GICs flowing in ground conductors have not been shown yet. The main goal of this work is to show, for the first time, with GIC data collected at a natural gas pipeline in southern Finland, how shock impact angles, combined with the pipeline's local time, affect the subsequent GIC enhancements. The remainder of the paper is organized as follows. Section 2 presents the data. Section 3 presents the results. The main results are discussed in section 4. Finally, Section 5 summarizes and concludes the article.

\section{Data}
	
	\subsection{Solar wind plasma and IMF data}
	
		In this work, we use the IP shock list provided by \cite{Oliveira2023c}. The list currently contains 603 events from January 1995 to May 2023. However, due to GIC data availability (see below), only 332 events from the list can be used in this study. The available events occurred from January 1999 to May 2023. Wind and ACE (Advanced Composition Explorer) solar wind plasma (particle number density, velocity, and temperature), and IMF data are used for shock detection and property computations. Solar wind data is explained by \cite{Ogilvie1995} for Wind, and by \cite{McComas1998} for ACE, whereas IMF data is detailed in \cite{Lepping1995} for Wind, and in \cite{Smith1998} for ACE. Before the computation of shock properties, the data was processed and interpolated as described in detail by \cite{Oliveira2023c}. \par

	\vspace{0.5cm}
	\subsection{Computation of shock impact angles and speeds}

		Shock properties including shock impact angles (\thxn) and shock speeds are computed with the Rankine-Hugoniot (RH) conditions. These conditions assume that energy and momentum across the shock front, normal magnetic field component, and tangential electric field component are conserved  \citep{Priest1981,Parks2004,Piel2010}. Such effects are computed with RH equations using solar wind velocity data, magnetic field data, and equations that combine both data sets. For example, a shock normal vector can be calculated with the equation \citep{Oliveira2017a,Oliveira2023c}:
			\begin{equation}
				\vec{n} = \pm\frac{\vec{B}_u \times (\vec{V}_d - \vec{V}_u) \times (\vec{B}_d - \vec{B}_u)}{|\vec{B}_u \times (\vec{V}_d - \vec{V}_u) \times (\vec{B}_d - \vec{B}_u)|}\,.
			\end{equation}

		In equation (1), $\vec{B}$ is the magnetic field vector, and $\vec{V}$ is the solar wind velocity vector. The indices $u$ and $d$ indicate observations in the upstream (non-shocked) and downstream (shocked) environments, respectively. Then, from a three-dimensional shock vector calculated with equation (1) and given by $\vec{n}=(n_x,n_y,n_z)$, \thxn{} can be computed as
			\begin{equation}
				\theta_{x_n} = \cos^{-1}(n_x)\,.
			\end{equation}
		
		We choose the (--) sign of $\vec{n}$ for \thxn{} defined in Geocentric Solar Ecliptic (GSE) coordinates because shock normals point toward the Earth in this case \citep{Schwartz1998}. Since the shocks are defined in GSE coordinates, a shock with \thxn{} = 180$^\circ$ indicates a purely frontal shock, whereas a shock with 90$^\circ$ $<$ \thxn{} $<$ 180$^\circ$ indicates an inclined shock, with the shock being more inclined as \thxn{} decreases. Figure 1 of \cite{Oliveira2023b} shows pictorial representations of a purely frontal shock and a highly inclined shock. Additionally, animations showing simulations of two shocks with different inclinations using real data can be found here: \url{https://dennyoliveira.weebly.com/phd.html}. \par

		According to the RH conditions, shock speeds of shocks with different inclinations can be calculated according to the expression \citep{Oliveira2017a,Oliveira2023c}:
			\begin{equation}
				v_s = \frac{(N_d\vec{V}_d - N_u\vec{V}_u) \cdot \vec{n}}{N_d - N_u}\,,
			\end{equation}
		where $N$ is the solar wind particle number density. As a result, the magnetosonic Mach number M$_s$ is computed as M$_s$ = $u_r/v_{ms}^f$, where $u_r = v_s - |\vec{V}_u|$, and $v_{ms}^f$ is the fast magnetosonic speed. A solar wind structure is classified as a true IP shock if M$_s$ $>$ 1. See section 2.3 of \cite{Oliveira2023c} for more details.

	\vspace{1.5cm}
	\subsection{Ground magnetometer data}

		Geomagnetic activity is represented by SuperMAG ground geomagnetic indices. SuperMAG data comprises of an array with hundreds of stations located worldwide for the computation of several geomagnetic indices to capture effects of magnetospheric and ionospheric currents at different latitudes \citep{Gjerloev2009}. Ring current effects are accounted for by the SuperMAG ring current SMR index \citep{Newell2012}, whereas auroral electrojet effects are represented by the SuperMAG total and regional SMU (upper envelope) and SML (lower envelope) indices \citep{Newell2012}. As detailed by \cite{Newell2012}, the SMR index is similar to the SYM-H index \citep{Iyemori1990}, but more low- and mid-latitude stations are used to compute the SMR index. Similar explanations are provided by \cite{Newell2011b}, who detail how the SMU and SML indices are calculated with more high-latitude stations in comparison to the traditional AU and AL indices \citep{Davis1966}. All SuperMAG data used in this study are 1-minute resolution data. \par 

		Local ground-based magnetic field response is represented by IMAGE (International Monitor for Auroral Geomagnetic Effects) data \citep{Viljanen1997}. IMAGE provides high-resolution data in northern Europe and eastern Greenland for studies of large-scale field-aligned current structures and dynamics of the high-latitude auroral electrojets \citep{Tanskanen2009}. IMAGE data resolutions are usually 10 s. In this study, we use data recorded at a single station, namely the Nurmij\"arvi (NUR) station, located in southern Finland at geographic coordinates 60.50$^\circ$ latitude and 24.65$^\circ$ longitude. The magnetic field components of the NUR data are represented in the north-ward direction (x component), eastward direction (y component), and downward direction (z component). \par

	\vspace{0.5cm}
	\subsection{GIC data from the Finnish natural gas pipeline} 

		GIC data recordings come from locations near the \man{} pipeline compression station in southern Finland (latitude 25.20$^\circ$, longitude 60.60$^\circ$) maintained by the Finnish Meteorological Institute and the European Community's Seventh Framework Programme. The \man{} GIC data is obtained by two magnetometers, one located at \man{}, and the other at NUR \citep{Pulkkinen2001a,Viljanen2010}. The ground \dbdt{} variations at \add{and auroral dynamics above} NUR account for natural variations of the geomagnetic field. The NUR field values are then subtracted from the \man{} field values, and the \change{different}{difference} is interpreted as field variations due to GIC effects \citep{Pulkkinen2001a,Viljanen2010}. Finally, by knowing the electromagnetic and geometric properties of the pipeline, along with the geoelectric field modeled by a framework shown in \cite{Pulkkinen2001b}, the actual measurements of GICs are determined. GICs measured at \man{} have an error of up to 1 A, which are smaller than the GIC peaks we intend to investigate in this work (GIC $>$ 5 A). Such GIC levels can cause overtime damage \citep{Beland2005,Gaunt2007,Rodger2017} and power disruptions in electric power transmission systems \citep{Allen1989,Bolduc2002,Oliveira2017d}. The \man{} GIC data and the NUR geomagnetic field data have both resolutions of 10 s. As recommended by \cite{Viljanen2010}, daily GIC average values were substrcted from the GIC data shown in this paper, even though these average values are very close to zero.

		\begin{figure}[t]
			\centering
			\includegraphics[width = 16cm]{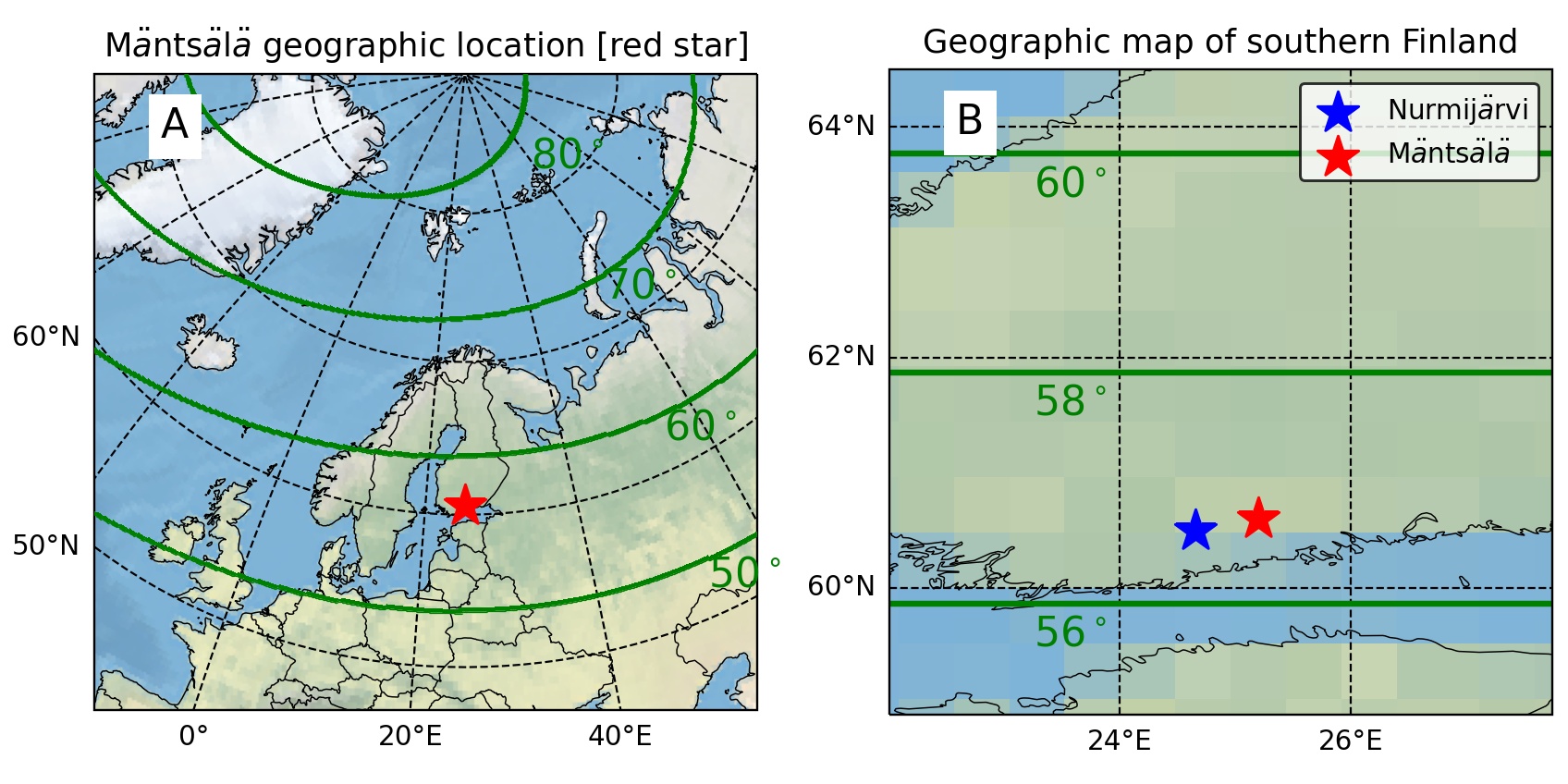}
			\caption{Panel A: Geographic position of the \man{} compression station in southern Finland (red star). The thick green lines are the magnetic latitudes from 50$^\circ$ in increments of 10$^\circ$ poleward. Panel B: snapshot of a southern Finland map showing the Nurmij\"arvi station (blue star), and the \man{} compression station (red star).}
			\label{mantsala_geo}
		\end{figure}

\section{Results}

	\subsection{Localizing geomagnetic field and GIC data in space and time}
	
\begin{figure}
			\centering
			\includegraphics[width = 14cm]{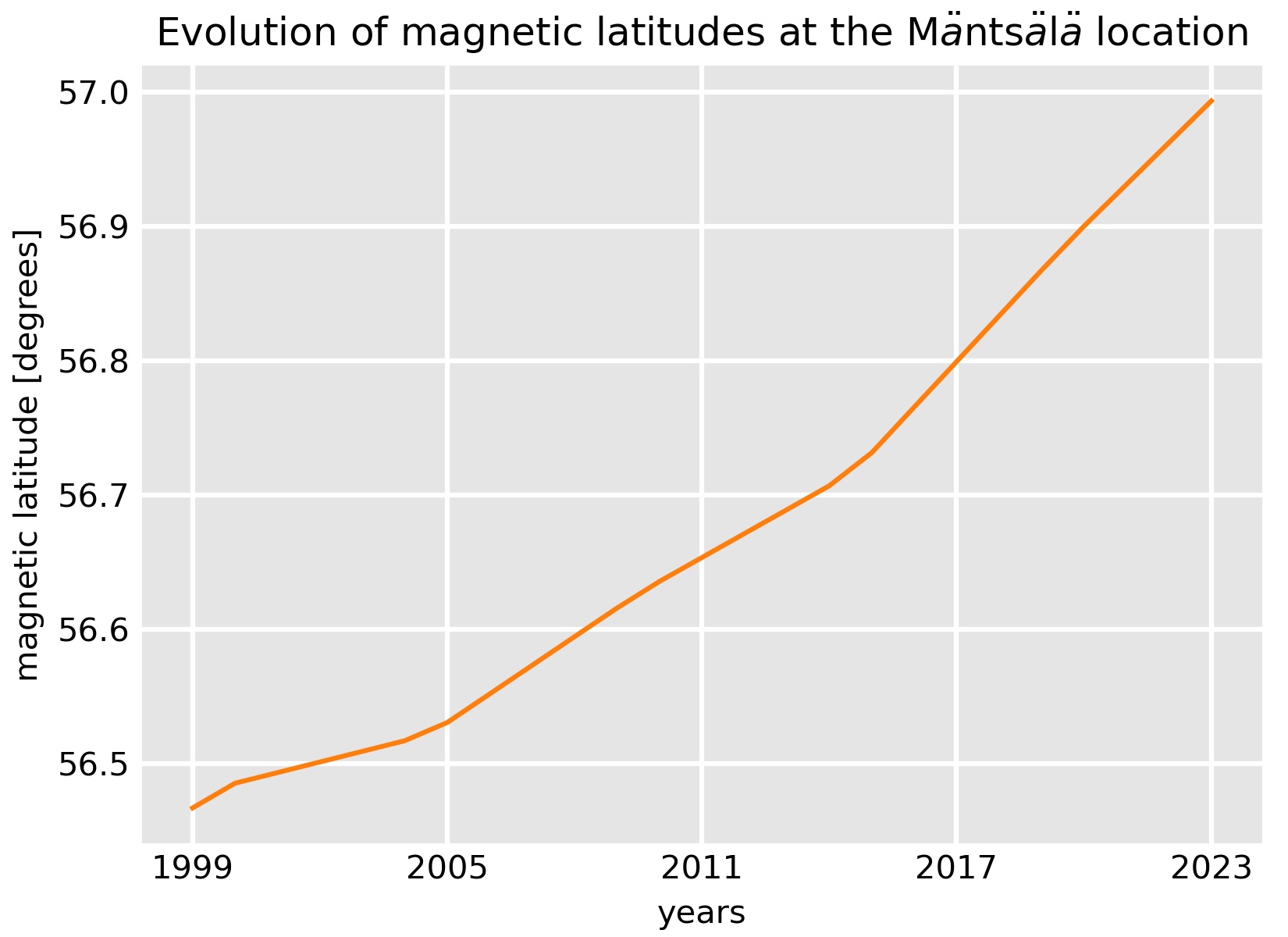}
			\caption{Time evolution of magnetic latitudes at \man{}'s geographic location from 1999 to 2023. This time span covers the full \man{} GIC pipeline system data set. The magnetic coordinates were calculated with the AACGM model.}
			\label{mantsala_mlats}
		\end{figure}

		Panel A of Figure \ref{mantsala_geo} shows the approximate location near the \man{} compression station in southern Finland. The thick green lines represent magnetic latitudes computed with the Altitude-Adjusted Corrected GeoMagnetic (AACGM) model \citep{Baker1989,Shepherd2014} for the year 2015. This figure shows that \man{} can certainly be underneath the auroral oval during intense geomagnetic storms and substorms, where it can reach magnetic latitudes as low as 50$^\circ$ \cite[e.g.,][]{Boteler2019,Hayakawa2020b}. Panel B in the same figure shows a snapshot of southern Finland with the geographic locations of \man{} (red star) and NUR (blue star). NUR is located approximately 40 km southwest of \man{}, which is way within the separation of 600 km between ground stations for adequate GIC modeling \citep{Ngwira2008}. The time evolution of \man{}'s magnetic latitude in the time span of this study (1999 to 2023) is shown in Figure \ref{mantsala_mlats}. The variation of the magnetic latitudes was near 0.5$^\circ$ in the period, which is neglegible for this study. Therefore, effects caused by different MLATs at \man{} can safely be ignored in this study. \par

\begin{figure}
			\centering
			\includegraphics[width = 14cm]{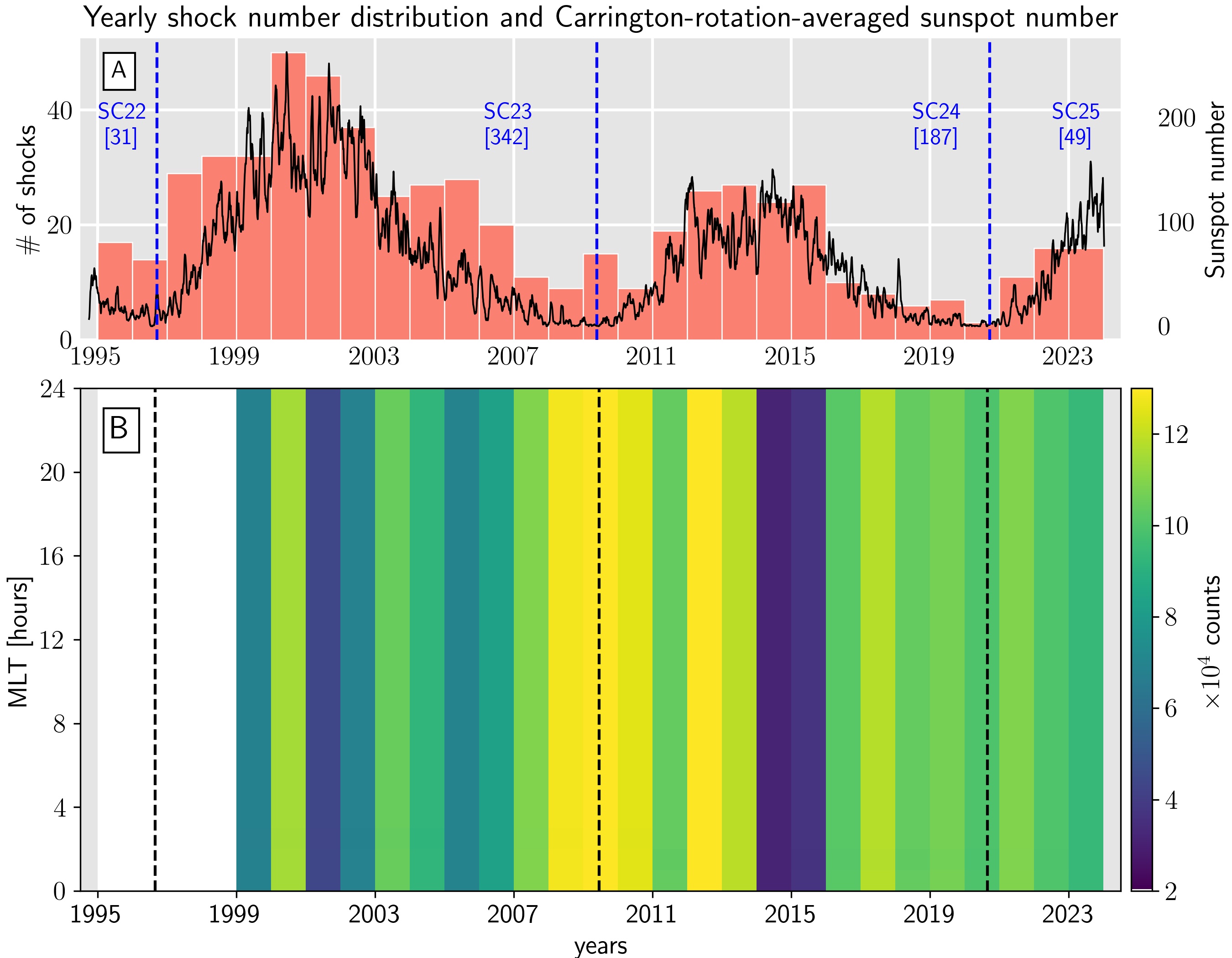}
			\caption{Panel A: Shock number distribution (salmon bars) and Carrington-rotation-averaged sunspot numbers (solid black line) from January 1995 to May 2023. This time span corresponds to the shock data base provided by \cite{Oliveira2023c}. Panel B: Number of GIC observations (counts) plotted as a function of year and magnetic local time (AACGM) at the \man{} natural gas pipeline system in the time span of the GIC data set (January 1999 to May 2023).}
			\label{ssn}
		\end{figure}

		Figure \ref{ssn} shows a comparison between the two main data sets used in this study, namely the shock and GIC data sets. Panel A shows annual number distributions of IP shocks in the \cite{Oliveira2023c} data base (salmon bars), and Carrington-averaged ($\sim27$ days) sunspot numbers from January 1995 to May 2023 (solid black line) \citep{Clette2015}. As discussed by \cite{Oliveira2023c}, both numbers of shocks and sunspots correlate quite well, meaning that shocks are more likely to occur during solar maxima \citep{Oh2002,Kilpua2015a,Oliveira2015a}. Panel B of Figure \ref{ssn} shows observation numbers of GIC observations plotted and color coded in 1 year $\times$ 1-hr MLT (AACGM magnetic local time) bins. The plot shows that there are no GIC measurements recorded at the \man{} pipeline before 1999. On the other hand, the number of observations in all MLT bins for a particular year are nearly the same, which indicates that eventual lack of observations may cover a few days. In addition, most years provide on average 8-10$\times10^4$ data points (1 data point $\equiv$ 10 s). Although there are good coverages for the maximum and declining phases of solar cycle 23 (SC23), there is good coverage for the ascending phase of SC24, but lower coverages during maximum and initial declining phases of SC24 (years 2015 and 2016). Finally, GIC data has good coverage in the ascending phase of the current SC25. Therefore, for this study, there's coverage of approximately two entire solar cycles with shocks and GIC concomitant data.

	\vspace{1cm}
	\subsection{Effects of \thxn{} on GICs: shock compression effects}

		In this subsection, we compare GIC effects caused by the impacts of two shocks with different inclinations, but with similar strengths as represented by magnetosonic Mach numbers. We choose a nearly frontal shock, hereafter NFS1, with \thxn{} = 161.77$^\circ$ and M$_s$ = 2.3, that struck the magnetosphere at 1400 UT on 18 April 2023. We also select a highly inclined shock, hereafter HFS1, with \thxn{} = 128.92$^\circ$ and M$_s$ = 2.6, that hit the magnetosphere at 1708 UT on 11 November 2004. This approach has been shown to be very effective for comparisons of geomagnetic activity triggered by shocks with different inclinations \citep{Wang2006a,Selvakumaran2017,Oliveira2018b,Shi2019b,Xu2020a}. As will become clearer later, these shocks were also chosen because their impacts on the magnetosphere occurred when \man{} was around dusk (MLT $\sim$16 hr and MLT $\sim19$ hr, respectively). In both plots, the black dashed vertical lines indicate the corresponding UT of shock impact of the magnetosphere. NFS1 was observed by Wind, whereas HIS1 was observed by ACE. Table \ref{table_compression} summarizes some general properties of the shocks used for comparisons of compression effects. \par

		\begin{table}
			\centering
			\begin{tabular}{c c c c c c c c}
				\hline
					Shock category & Date 	&	UT$^\dagger$ 	& MLT$^{\dagger\dagger}$ [hr]	&	\thxn{} [$^\circ$]	& $v_s$ [km/s] & $DP_d/DP_u$$^{\dagger\dagger\dagger}$	&	M$_s$ \\
				\hline
					NFS1  & 2023/04/18 	& 	1400  &	 15.9	& 	161.77 	&	545.83 	& 4.67	&	2.3 \\
					HIS1 & 2004/11/11  &	1708  &	 19.2	&   128.97	& 	495.19 	& 2.48	&	2.6 \\
				\hline
					\multicolumn{7}{l}{$^\dagger$ UT of shock impact on the magnetosphere.} \\
					\multicolumn{7}{l}{$^{\dagger\dagger}$ \man's MLT at UT of shock impact on the magnetosphere.} \\
					\multicolumn{7}{l}{$^{\dagger\dagger\dagger}$ Downstream to upstream dynamic pressure ratio: $DP_d/DP_u$ = $N_dV_d^2 / N_u V_u^2$.}
			\end{tabular}
			\caption{Comparison of parameters for a nearly frontal shock and a highly inclined shock. The focus is on shock compression effects.}
			\label{table_compression}
		\end{table}

		\begin{figure}
			\centering
			\includegraphics[width = 16cm]{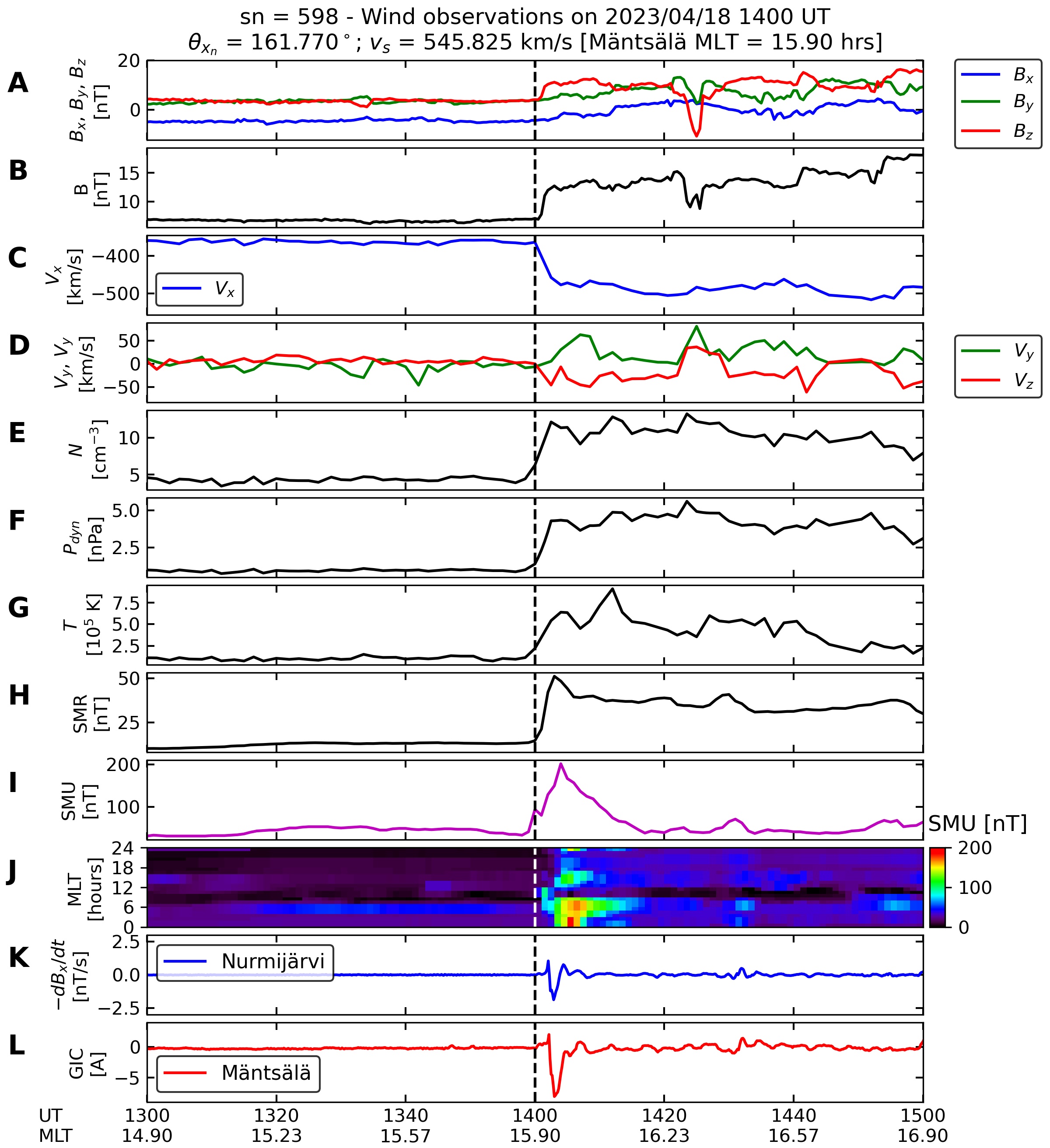}
			\caption{Solar wind and IMF, geomagnetic index, and \man{} GIC data for the shock event of 18 April 2023. Solar wind and IMF data were recorded by ACE in this case. Note that NUR ground magnetic field variations are plotted as --\dbxdt{} to follow GIC trends at \man{}.}
			\label{mantsala_598}
		\end{figure}

\begin{figure}[t]
			\centering
			\includegraphics[width = 16cm]{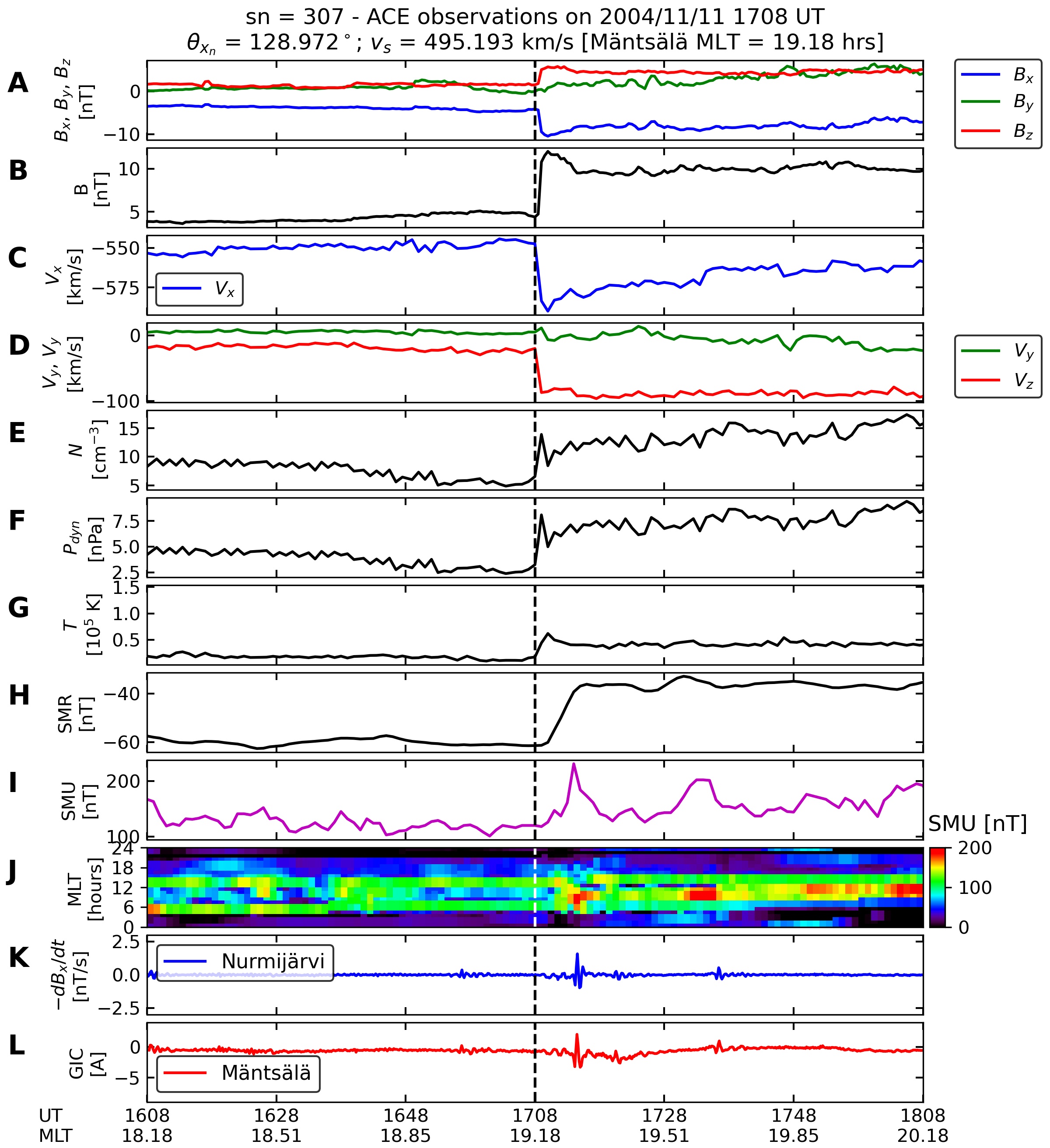}
			\caption{Solar wind and IMF, geomagnetic index, and \man{} GIC data for the shock event of 11 November 2004. Solar wind and IMF data were recorded by ACE in this case. Note that NUR ground magnetic field variations are plotted as --\dbxdt{} to follow GIC trends at \man{}.}
			\label{mantsala_307}
		\end{figure}

		Figures \ref{mantsala_598} and \ref{mantsala_307} show time series for solar wind, IMF, geomagnetic index, ground magnetic field, and GIC data for the NFS1 and HFS1, respectively. Both figures show the three components of the IMF (panels A), IMF magnitude (panel B), three components of the solar wind velocity (x, panel C; y and z, panel D), plasma number density (panel E), dynamic pressure $m_p N V^2$, where $m_p$ is the proton mass (panel F), temperature (panel G), SMR (panel H), SMU (panel I), regional SMU plotted as a function of time and MLT (panel J), ground \dbxdt{} at NUR (panel K); and GIC at \man{} (panel L). In order to compare geomagnetic effects caused by IP shock compression, we plot data $\pm$1 hr around shock impact time (dashed black vertical line) and focus on the following 20 minutes. This time has been shown to be adequate to focus only on shock compression effects \citep{Selvakumaran2017,Oliveira2018b,Rudd2019}. \par

		Both figures show that IMF conditions are very similar before shock impacts (B$_x$ slightly negative, with B$_y$ and B$_z$ near zero values). After shock impacts, the IMF magnitudes increase from values near 5 nT to values around 13 nT. In both time series, V$_x$ shows a sharper decrease, with V$_z$ showing a much more intense variation in the HIS1 case with respect to the NFS1 case. Although the shock compression rates (downstream to upstream plasma number density ratio) are similar (2.6 and 2.2, respectively), the dynamic pressure compression ratio is higher in the case of the NFS1 (see Table \ref{table_compression}). These observations indicate that the HIS1 is indeed more inclined and stronger than NFS1. As theoretically demonstrated by \cite{Samsonov2011a}, nearly frontal shocks compress the magnetosphere mostly in the x direction, whereas highly inclined shocks compress the magnetosphere more significantly in the y and z directions in comparison to y and z directions in nearly frontal shocks. As demonstrated by many works \citep[see, e.g., ][]{Oliveira2023b}, these asymmetric compression effects caused by highly inclined shocks usually lead to very different geomagnetic activity in terms of asymmetries and intensities in comparison to nearly frontal shocks. \par

		The shock impact angle effects caused by the two shocks can be clearly seen in the remaining panels of Figures \ref{mantsala_598} and \ref{mantsala_307}. Panels H and I show that SMR and SMU are more intense and develop faster in the NFS1 case in comparison to the HIS1 case. These effects have been shown in many works, including simulations and observations \citep{Takeuchi2002b,Guo2005,Wang2006a,Rudd2019,Shi2019b,Oliveira2021b}. The regional SMU index shows strong enhancements around MLT = 6 hours, but a relative SMU change is higher in the NFS1 case. Ground --\dbxdt{} variations at NUR are more intense (magnitude near 2.5 nT/s) and peak earlier in the first case in comparison to the second case. These effects have already been reported to be observed in ground magnetometer data in geospace \citep{Oliveira2020d} and on the ground \citep{Takeuchi2002b,Oliveira2020d,Oliveira2021b}. As expected, similar trends are observed in the GIC observations shown by the red lines in both plots, which are supported by the papers mentioned above and many others \citep{Oliveira2018a,Oliveira2023b}. Finally, our NUR --\dbxdt{} and \man{} GIC observations agree with a trend reported by \cite{Viljanen2010}, who pointed out that GIC measurements almost always follow --\dbxdt{} measurements (which are plotted in the figures) in the x direction because the \man{} pipeline is positive-oriented in the eastward direction from \man.

	\vspace{0.5cm}
	\subsection{Effects of \thxn{} on GICs: substorm effects}

\begin{figure}
			\centering
			\includegraphics[width = 16cm]{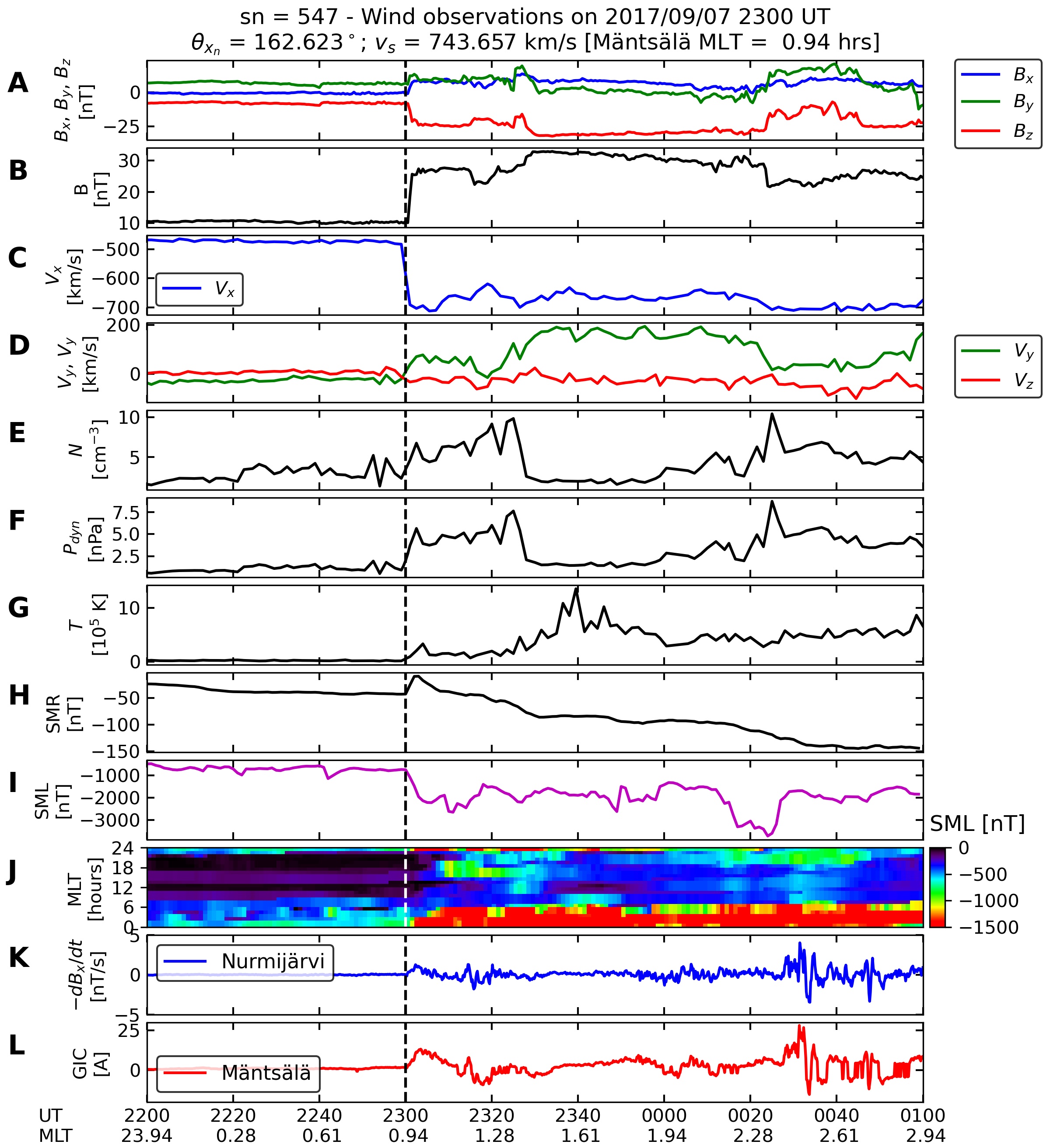}
			\caption{Solar wind and IMF, geomagnetic index, and \man{} GIC data for the shock event of 7 September 2017. Solar wind and IMF data were recorded by ACE in this case. Note that NUR ground magnetic field variations are plotted as --\dbxdt{} to follow GIC trends at \man{}.}
			\label{mantsala_547}
		\end{figure}

\begin{figure}
			\centering
			\includegraphics[width = 16cm]{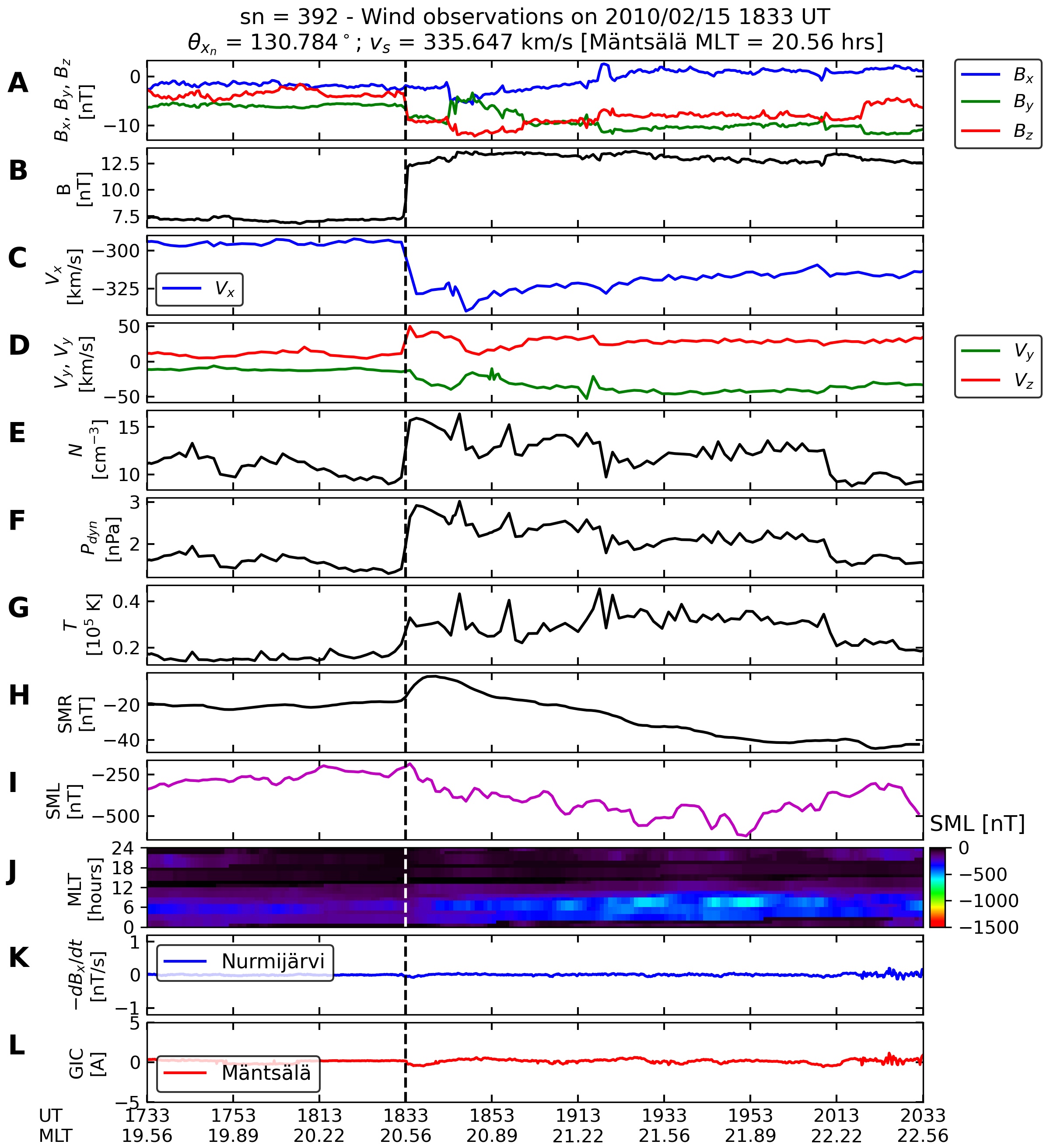}
			\caption{Solar wind and IMF, geomagnetic index, and \man{} GIC data for the shock event of 15 February 2010. Solar wind and IMF data were recorded by ACE in this case. Note that NUR ground magnetic field variations are plotted as --\dbxdt{} to follow GIC trends at \man{}.}
			\label{mantsala_392}
		\end{figure}

		Now we focus on GIC enhancements caused by substorm effects triggered by shocks. We select two shocks, a nearly frontal shock (NFS2), with \thxn{} = 162.62$^\circ$ and M$_s$ = 2.4, and a highly inclined shock (HFS2), with \thxn{} = 130.78$^\circ$ and M$_s$ = 2.4. The impact of the NFS2 on the magnetosphere occurred at 2300 UT of 07 September 2017, whereas the impact of the HFS2 took place at 1833 UT of 15 February 2010. Table \ref{table_substorm} summarizes the main properties of NFS2 and HIS2. These shocks were selected because their impacts occurred at MLT = 0.9 hr and MLT = 20.6 hr \add{at \man} for NFS2 and HIS2, which allows for the observation of magnetotail effects on GICs around the magnetic midnight terminator (MLT = 00 hr). Solar wind, IMF, geomagnetic index, ground magnetic field, and GIC data are plotted for the NFS2 and HFS2 in Figures \ref{mantsala_547} and \ref{mantsala_392}. These figures are similar to Figures \ref{mantsala_598} and \ref{mantsala_307}, except for two differences: first, the SMU index time series and regional SMU index are replaced by SML, and data is plotted 1 and 2 hrs around the shock impact onset. The time span after the shock impacts are adequate to account for substorm activity triggered by solar wind driving including shocks \citep{Bargatze1985,Freeman2004,Oliveira2014b,Oliveira2024a}. \par

			\begin{table}
				\centering
				\begin{tabular}{c c c c c c c c}
					\hline
						Shock category & Date 	&	UT$^\dagger$ 	& MLT$^{\dagger\dagger}$ [hr]	&	\thxn{} [$^\circ$]	&	$v_s$ [km/s]	&	$DP_d/DP_u$$^{\dagger\dagger\dagger}$ &	M$_s$ \\
					\hline
						NFS2  & 2017/09/07 	& 	2300  &	 0.94	& 	162.62 	&	743.66 	& 4.27	& 	2.4 \\
						HFS2  & 2010/02/15  &	1833  &	 20.6	&   130.78	& 	335.65 	& 1.90	& 	2.4 \\
					\hline
					\multicolumn{7}{l}{$^\dagger$   UT of shock impact on the magnetosphere.} \\
					\multicolumn{7}{l}{$^{\dagger\dagger}$  \man's MLT at UT of shock impact on the magnetosphere.} \\
					\multicolumn{7}{l}{$^{\dagger\dagger\dagger}$ Downstream to upstream dynamic pressure ratio: $DP_d/DP_u$ = $N_dV_d^2 / N_u V_u^2$.}
				\end{tabular}
				\caption{Comparison of parameters for a nearly frontal shock and a highly inclined shock. The focus is on effects caused by shock-triggered substorms.}
				\label{table_substorm}
			\end{table}

		In both shock cases, IMF B$_z$ values in the upstream region were close to --5 nT. Preconditioning effects are very important conditions for substorm triggering \citep{Zhou2001,Yue2010}, and they determine whether a substorm is triggered or not. On the other hand, it is very clear that IMF B$_z$ is much more depleted in the NSF2 case in comparison to the HFS2 case because the nearly head-on impact amplifies the southward condition of IMF B$_z$ in comparison to the other case. These different magnetospheric compression conditions were shown with simulation by \cite{Oliveira2014b} to be very effective in determining the intensity of substorm triggering, being much more intense in the frontal case in comparison to the inclined case. This is clearly seen in Table \ref{table_substorm}, with the downstream to upstream ram pressure ratio being higher in the NFS2 case in comparison to the HIS2 case. SMR amplitudes are more intense and occur earlier in the nearly frontal case in comparison to the highly inclined case \citep{Guo2005,Selvakumaran2017,Rudd2019}. Time series for the SML index (magenta lines) indicate much more intense magnetotail activity in the first case (SML $<$ --3,000 nT) in comparison to the second case (SML around --600 nT). Similar effects were shown with SuperMAG data by \cite{Oliveira2015a}. The regional SuperMAG index data show very intense SML activity (SML $<$ --1,500 nT) around the magnetic midnight for the NFS2, whereas mild SML activity (SML $\sim$ --500 nT) is seen near dawn for the HIS2. These results agree with the simulations conducted by \cite{Oliveira2014b} for shocks with different orientations. \par

		\begin{figure}
			\centering
			\includegraphics[width = 16cm]{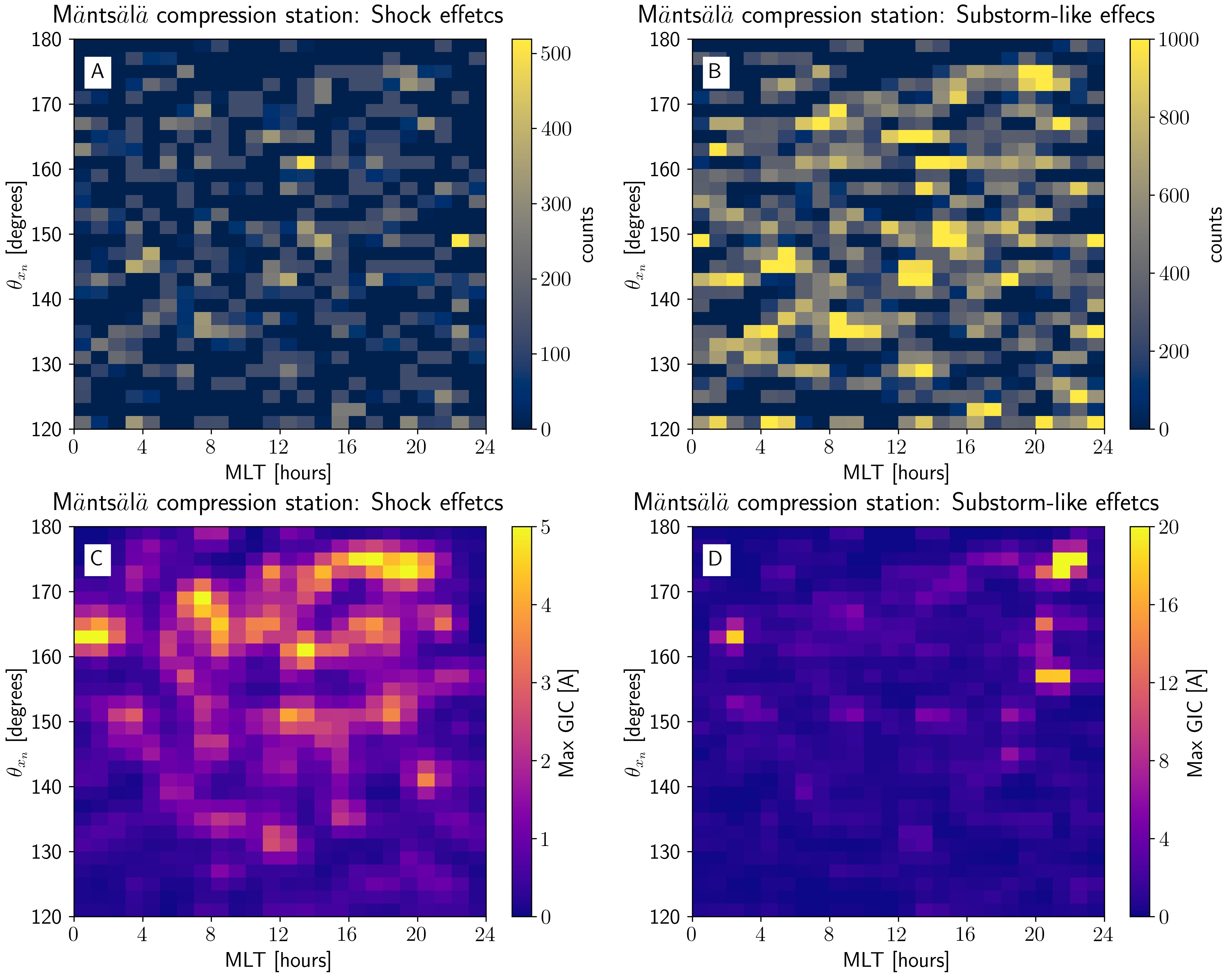}
			\caption{Superposed epoch analysis of GIC response recorded at the \man{} compression station during shock compression effects (first 20 minutes after shock impat, first column) and during substorm-like effects (within 20 and 120 minutes after shock impact, second column). Color-codes represent observation counts (panels A and B), and GIC peaks (panels C and D). All data are plotted in MLT $\times$ \thxn{} bins.}
			\label{mantsala_hist}
		\end{figure}

		As depicted in Figures \ref{mantsala_547} and \ref{mantsala_392}, after shock impacts, some NUR --\dbxdt{} variations and \man{} GIC variations are observed at their respective locations in the NFS2 case, but close to none observations are seen after shock impact in the HIS2 case (note that panels K and L in the figures are not to scale). Later, intense ground --\dbxdt{} and GIC variations are seen around 90 minutes after shock impacts in the NFS2 case, whereas no noticeable observations are recorded at NUR and \man. These observations strongly agree with the results shown by \cite{Oliveira2021b}: ground --\dbdt{} variations are more intense during substorms triggered by nearly head-on shock impacts on the magnetosphere. These results are also supported by the statistical and superposed epoch analysis study reported by \cite{Oliveira2024a}.

	\vspace{0.5cm}
	\subsection{Statistical results}

		A superposed epoch analysis of GIC peaks for all shocks is shown in Figure \ref{mantsala_hist}. The top panels show counts or number of data points or observations (1 data point $\equiv$ 10 s) of GIC peaks caused by shock compressions (panel A) and substorm effects (panel B). The lower panels show the GIC peaks associated with shock compression effects (panel C) and substorm effects (panel D). In all panels, data are plotted as a function of MLT and \thxn, with the color codes representing number of observations (panels A and B) and GIC peaks (panels C and D). \par

		\begin{figure}
			\centering
			\includegraphics[width = 13cm]{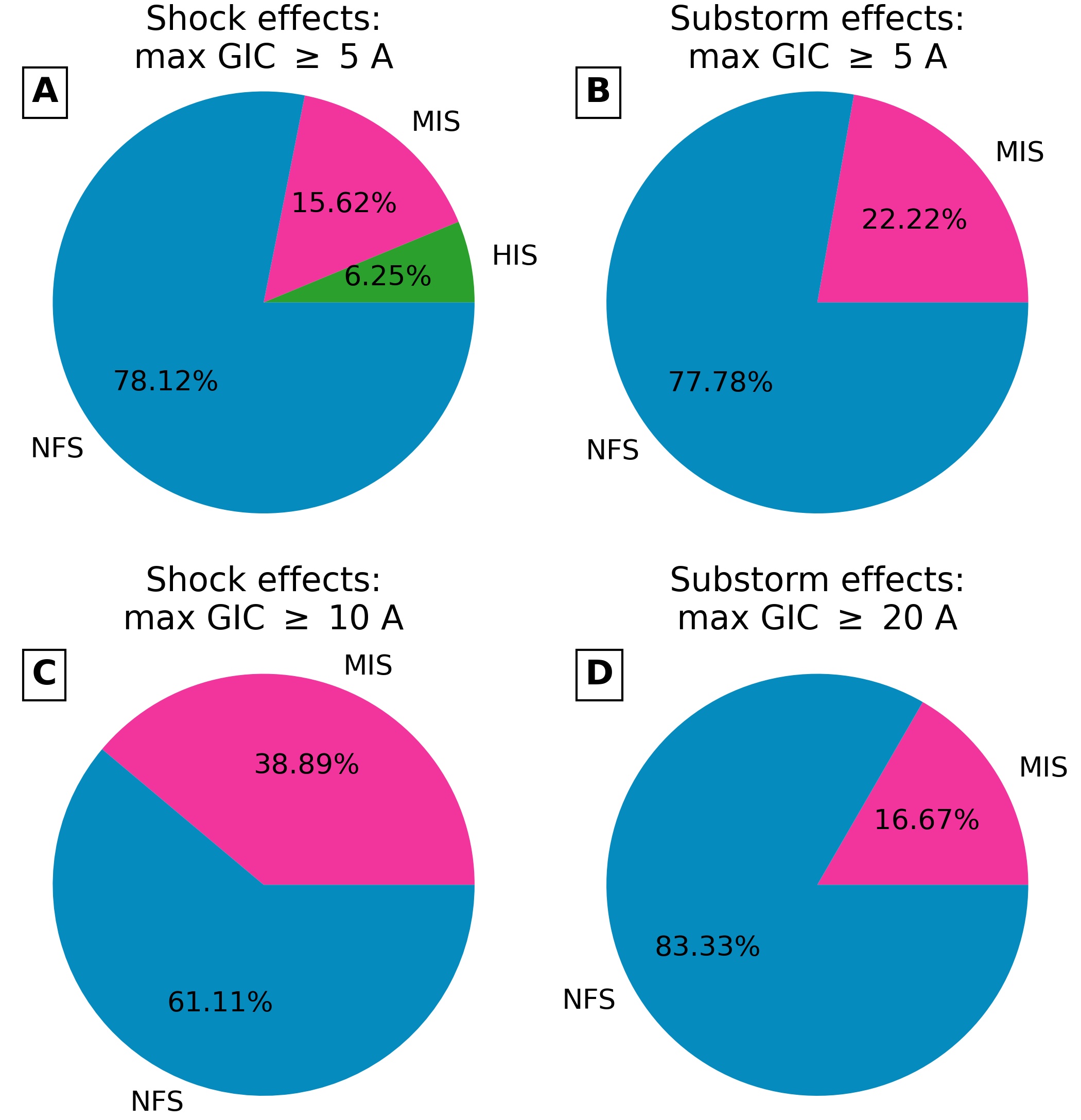}
			\caption{Pie diagrams documenting relative GIC peak response $\geq$ 5 A for shock effects (A), and substorm effects (B). GIC peaks $\geq$ 10 A are shown in panel C (shock effects), whereas GIC peaks ($\geq$ 20 A) resulting from magnetotail activity are shown in panel D. Blue colors indicate NFS; magenta colors, MIS; and green colors, HIS.}
			\label{manpie}
		\end{figure}

		In order to explore shock impact angle effects on the subsequent GIC peak response, we classify events as highly inclined shocks (HIS, with \thxn{} $<$ 140$^\circ$); moderately inclined shocks (MIS, 140$^\circ$ $\leq$ \thxn $<$ 160$^\circ$); and nearly frontal shocks (NFS, \thxn $\geq$ 160$^\circ$). Such shock impact angle classifications have been shown to be very effective in capturing impact angle effects on the following geomagnetic activity caused by shocks \citep[e.g., ][]{Wang2006a,Oliveira2015a,Selvakumaran2017,Oliveira2018b,Rudd2019,Shi2019b,Xu2020a,Oliveira2024a}. \par

		\begin{table}
		\centering
		\begin{tabular}{c r r r r r r r r r}
			\hline
				Shock 		&  Shock  &    MLT1   &	  \thxn{}	  &  $v_s$  & $\Delta$SMR & max &  max       &	 MLT2     & \# \\
				Date		&	UT    & 	[h]   &    [$^\circ$] &  [km/s] &  [nT] &  SMU [nT] &  GIC [A]   &   [h] 	  &      \\  
			\hline
				2000/06/08  &  0911   &    11.26   &    173.48  &  882.01  &   60.05   &    699.0   &   12.23     &   11.26     &       4 \\
				2000/10/04  &  1359   &    16.06   &    175.66  &  480.01  &   44.95   &    688.0   &    9.38     &   16.27     &      16 \\
				2000/11/10  &  0627   &     8.52   &    165.01  &  879.51  &   73.79   &    491.0   &   10.29     &    8.55     &       4 \\
				2001/03/31  &  0052   &     2.94   &    150.19  &  672.23  &  156.40   &    514.0   &    7.71     &    3.09     &       5 \\
				2001/04/04  &  1453   &    16.95   &    165.92  &  890.43  &   51.78   &    634.0   &    6.80     &   16.99     &       2 \\
				2001/04/11  &  1548   &    17.87   &    151.57  &  709.69  &   33.11   &    687.0   &    8.16     &   18.19     &       5 \\
				2001/04/28  &  0458   &     7.04   &    168.47  &  923.43  &   58.26   &    779.0   &   12.79     &    7.09     &       9 \\
				2001/10/11  &  1658   &    19.03   &    175.28  &  567.15  &   60.11   &    409.0   &    5.08     &   19.08     &       1 \\
				2001/10/21  &  1647   &    18.85   &    172.91  &  627.37  &   71.10   &    593.0   &    6.21     &   18.88     &       5 \\
				2002/03/18  &  1321   &    15.41   &    174.82  &  542.50  &   76.53   &    405.0   &    7.67     &   15.45     &       2 \\
				2002/04/17  &  1106   &    13.16   &    161.71  &  440.03  &   53.64   &    662.0   &   18.96     &   13.16     &       9 \\
				2002/04/19  &  0834   &    10.63   &    165.84  &  753.69  &   34.60   &    325.0   &    5.55     &   10.65     &       1 \\
				2002/04/23  &  0448   &     6.86   &    169.57  &  686.23  &   54.66   &    450.0   &    5.06     &    6.88     &       1 \\
				2002/05/23  &  1049   &    12.88   &    150.31  &  710.92  &   82.52   &   1742.0   &   10.46     &   12.91     &       1 \\
				2003/05/29  &  1859   &    21.04   &    164.65  &  938.49  &   54.24   &    761.0   &    9.81     &   21.26     &      35 \\
				2003/10/24  &  1523   &    17.44   &    175.14  &  652.94  &   59.77   &   1107.0   &   12.07     &   17.70     &      34 \\
				2003/11/04  &  0625   &     8.47   &    161.21  &  842.14  &   68.94   &    734.0   &    6.96     &    8.47     &       4 \\
				2003/11/20  &  0802   &    10.09   &    163.02  &  684.54  &   34.80   &    483.0   &    7.34     &   10.10     &       2 \\
				2004/11/07  &  1827   &    20.50   &    141.12  &  601.60  &  111.21   &    585.0   &   11.39     &   20.53     &       5 \\
				2004/11/09  &  0928   &    11.51   &    131.34  &  805.34  &   30.19   &    546.0   &    5.23     &   11.56     &       1 \\
				2004/11/09  &  1848   &    20.85   &    174.12  &  880.53  &   68.26   &    770.0   &    5.49     &   20.88     &       2 \\
				2005/01/21  &  1711   &    19.23   &    172.56  & 1070.10  &   99.92   &    615.0   &   13.34     &   19.25     &       5 \\
				2011/09/26  &  1235   &    14.58   &    172.77  &  527.15  &   52.07   &    283.0   &    8.98     &   14.58     &       7 \\
				2012/01/24  &  1502   &    17.02   &    160.39  &  496.07  &   51.29   &    634.0   &    7.20     &   17.06     &       2 \\
				2012/03/08  &  1103   &    13.04   &    168.50  &  958.72  &   76.15   &    528.0   &    6.27     &   13.31     &       4 \\
				2012/03/12  &  0914   &    11.22   &    165.87  &  524.25  &   62.50   &    554.0   &    6.56     &   11.27     &       3 \\
				2012/07/14  &  1807   &    20.10   &    173.31  &  667.09  &   55.90   &    590.0   &    5.80     &   20.19     &       5 \\
				2012/09/03  &  1213   &    14.20   &    172.94  &  425.96  &   50.59   &    480.0   &    6.21     &   14.25     &      11 \\
				2017/09/07  &  2300   &     0.94   &    162.62  &  743.66  &   29.21   &    936.0   &   13.30     &    0.99     &      71 \\
				2023/03/15  &  0427   &     6.35   &    157.42  &  608.24  &   50.22   &    418.0   &    7.89     &    6.38     &       5 \\
				2023/04/18  &  1359   &    15.88   &    161.77  &  545.83  &   38.10   &    202.0   &    8.09     &   15.93     &       7 \\
				2023/05/08  &  1356   &    15.83   &    135.55  &  441.75  &   28.74   &    258.0   &    5.06     &   15.90     &       1 \\
			\hline
		\end{tabular}
		\caption{Table for shock properties and the subsequent geomagnetic index/GIC peak responses. In the table, MLT1 indicates \man{} MLT at shock impact; $v_s$, shock speed; $\Delta$SMR, SMR index variation (SMR peak after shock compression minus background); MLT2, \man{} MLT at GIC peak occurrence; and the rightmost column indicates the number of GIC peaks greater than 5 A within 20 minutes of shock impact for each event.}
		\label{table1}
	\end{table}

		In the shock compression case, panels show that observation counts indicate that most bins show observation numbers less than 200, but a few bins show observation numbers greater than 300. In the other case, most bins indicate observation numbers greater than 600, and fewer bins indicate observation numbers greater than 800. Therefore, although the numbers of observations are spread out in the bins, there are no particular biases introduced by either MLT or \thxn{} in the observations. The overall number of observations is higher in the substorm effects case in comparison to the shock compression case due to the time span of observations around shock onset (within 20 minutes and within 20 and 120 minutes for both cases, respectively). \par

		GIC peaks ($>$ 5 A) in the shock compression case (Figure \ref{mantsala_hist}C) occur more often for NFSs. There are very few peaks (1-2 A) for the HIS case, but they are apparently more concentrated around MLT = 12 hr. MISs show peak intensities in between the HIS and NFS categories, with MLT coverage in between. Although these observations agree with previous works \citep{Oliveira2018b,Xu2020a}, our results are novel in two ways: first, this is the first time IP shock impact angle effects are observed on actual GICs, and, second, there is broader and more intense GIC \change{peask}{peak} response around the dusk sector (MLT $\sim$ 18 hr), which is much more evident for NFSs. \par

		\begin{table}
		\centering
		\begin{tabular}{c r r r r r r r r r}
			\hline
				Shock 		&  Shock  &    MLT1   &	  \thxn{}	  &  $v_s$  & min SMR & min &  max       &	 MLT2     & \# \\
				Date		&	UT    & 	[h]   &    [$^\circ$] &  [km/s] &  [nT] &  SML [nT] &  GIC [A]   &   [h] 	  &      \\   
			\hline
				2000/04/06  &  1640   &    18.74   &    165.48  &  730.57  &  --65.97   &   --2367   &   20.28     &   20.24     &      35 \\
				2000/10/04  &  1359   &    16.06   &    175.66  &  480.01  &  --89.23   &   --1433   &    6.17     &   16.47     &       5 \\
				2001/04/11  &  1518   &    17.37   &    144.48  &  683.44  &   --2.83   &   --2920   &    9.20     &   18.47     &      67 \\
				2001/04/11  &  1548   &    17.87   &    151.57  &  709.69  &  --92.45   &   --2920   &    9.20     &   18.47     &      63 \\
				2001/10/21  &  1647   &    18.85   &    172.91  &  627.37  &  --19.85   &   --1565   &    7.19     &   20.07     &      15 \\
				2002/05/23  &  1049   &    12.88   &    150.31  &  710.92  &  --66.38   &   --1076   &    7.42     &   13.91     &      12 \\
				2002/09/07  &  1636   &    18.66   &    161.21  &  603.65  &  --140.48   &   --2285   &    9.32     &   20.03     &      19 \\
				2003/05/29  &  1859   &    21.04   &    164.65  &  938.49  &  --69.10   &   --2455   &    6.11     &   22.89     &      11 \\
				2003/10/24  &  1523   &    17.44   &    175.14  &  652.94  &  --46.97   &   --2072   &    8.73     &   18.00     &      10 \\
				2004/11/09  &  1848   &    20.85   &    174.12  &  880.53  &  --228.45   &   --2395   &   42.82     &   21.88     &     268 \\
				2005/01/21  &  1711   &    19.23   &    172.56  & 1070.10  &   --2.75   &   --4054   &   27.44     &   21.20     &      84 \\
				2012/03/08  &  1103   &    13.04   &    168.50  &  958.72  &    2.85   &   --551   &    5.46     &   13.42     &       1 \\
				2012/09/30  &  2306   &     1.09   &    153.04  &  443.27  &  --100.80   &   --1152   &    7.17     &    2.99     &       9 \\
				2017/08/31  &  0539   &     7.59   &    150.47  &  411.36  &   30.28   &   --273   &    6.70     &    9.09     &       1 \\
				2017/08/31  &  0538   &     7.58   &    167.13  &  433.82  &   26.88   &   --273   &    6.70     &    9.09     &       1 \\
				2017/09/07  &  2300   &     0.94   &    162.62  &  743.66  &  --144.88   &   --3709   &   28.18     &    2.46     &     233 \\
				2023/02/26  &  1922   &    21.27   &    157.57  &  555.82  &  --27.30   &   --903   &    5.93     &   23.00     &       5 \\
				2023/04/23  &  1735   &    19.48   &    156.04  &  569.47  &  --115.57   &   --1608   &   27.32     &   20.95     &     183 \\
			\hline
		\end{tabular}
		\caption{Table for substorm effects and the subsequent geomagnetic index/GIC peak responses. This table is similar to Table \ref{table1}, but with two modification: $\Delta$SMR and maximum SMU are now replaced by minimum SMR and minimum SML values, respectively. The time of interest is in between 20 minutes and 120 minutes after shock impacts.}
		\label{table2}
	\end{table}

	In Figure \ref{mantsala_hist}D, GIC peaks are shown in the same way as in panel C, but for the case accounting for substorm effects. However, since GIC peaks in this case are more intense, the panel highlights GIC peaks greater than 10 A. Most of these peaks occur around MLT = 00 hr for NFSs, but a few peaks occur for MISs with 20 hr $<$ MLT $<$ 22 hr. These results agree withVsubstorm effects triggered by nearly head-on shock impacts on the subsequent ground \dbdt{} variations in the case study reported by \cite{Oliveira2021b}, and the statistical study provided by \cite{Oliveira2024a}. \par

	Figure \ref{manpie} shows the same data represented in panels C and D of Figure \ref{mantsala_hist}, but organized  in pie diagrams with the relative occurrence number of events with GIC peaks greater than 5 A (panels A and B), and GIC peaks greater than 10 A (panel C) and greater than 20 A (panel D). Shock inclination categories are represented in blue, NFS; magenta, MIS; and green, HIS. The first column is for events caused by magnetospheric compression by the shocks (within 20 minutes after shock onset), whereas the second column is for magnetotail or substorm effects (between 20 minutes and 120 minutes after shock onset).  \par

	Results show that NFSs dominate GIC peaks for both GIC thresholds and space weather drivers (magnetospheric compression and substorm effects). For GIC peaks $\geq$ 5 A, compression effects, NFSs account for more than three quarters of the events, with $\sim$16\% of events classified as MISs, and $\sim$6\% classified as HISs. As shown in Table \ref{table1}, there is only one event classified as HIS. Still for compression effects, nearly three quarters of the events are NFS, whereas nearly one quarter of the events are MIS. There are no HIS events. Therefore, these results clearly show that shock impact angles significantly control the subsequent GIC peaks at \man, particularly for GIC peaks greater than 20 A occurring during substorm events (panel D). Tables \ref{table1} (shock compression effects) and \ref{table2} (substorm effects) summarize the shock properties, geomagnetic index, and GIC peak responses to all events investigated in this study with GIC peaks grater than 5 A.

\section{Discussion}
	
	In this investigation, we used the shock data base provided by \cite{Oliveira2023c} and GIC data measurements from the Finnish natural gas pipeline system to study shock impact angle effects on the subsequent GICs. Our observations are based on previous studies of shock impact angle effects on ground \dbdt{} variations, which are the space weather drivers of GICs. For example, \cite{Oliveira2018b} showed that nearly frontal shocks trigger more intense ground \dbdt{} variations following the shock impact in comparison to highly inclined shocks at all latitudes. In another work, \cite{Oliveira2021b} demonstrated with a case study that a nearly frontal shock triggered a super substorm (SML $<$ --2,500 nT), and a highly inclined shock triggered an intense substorm (--2,500 nT $<$ SML $\leq$ --2,000 nT), even though both shocks had similar strengths. As a result, ground \dbdt{} variations were more intense, occurred earlier, and covered larger geographic areas in North America and western Greenland in the first case in comparison to the second case. All these results suggest that nearly frontal shocks compress the magnetosphere more symmetrically in comparison to highly inclined shocks, with the former enhancing more effectively the most important current systems in the magnetosphere-ionosphere system leading to higher geomagnetic activity \citep{Takeuchi2002b,Guo2005,Wang2006a,Oliveira2014b,Oliveira2015a,Samsonov2015,Selvakumaran2017,Oliveira2018a,Shi2019b,Xu2020a,Oliveira2023b}. \par

	The statistical study of \cite{Oliveira2024a} confirmed the case study of \cite{Oliveira2021b} by showing that ground \dbdt{} variations induced by nearly frontal shocks (i) are more intense, (ii) cover larger geographic areas including (iii) more equator-ward regions in comparison to highly inclined shocks. However, as outlined in the introductory section, though ground \dbdt{} variations are considered the space weather drivers of GICs, actual GIC effects can only be adequately quantified with the use of ground conductivity models \citep{Beggan2015,Espinosa2019,Wang2021}. This is a difficult task because GICs present a spectral dependence on ground \dbdt{} variations due to their interaction with the non-uniform Earth's conductivity structures in many layers whose modeling is quite complex \citep{Gannon2017,Kelbert2020,Juusola2020}. In addition, the geometry of the conductors of interest must be known for accurate computation of GICs. However, since this has been accomplished with the \man{} GIC data set, we were able to investigate shock impact angle effects on actual GIC measurements for the first time. Although GICs were tackled on in very similar magnetic latitudes, we can clearly conclude from our results that the more frontal the shock, the more intense the GIC amplitudes during compressions after the shocks (GIC $>$ 5 A), and during magnetotail energy inputs due to energetic particle injections during substorm times (GIC $>$ 20). These results build upon the work of \cite{Tsurutani2021} who observed very intense storm-time GIC peaks ($>$ 30 A), but our contribution indicates that intense GICs ($>$ 5 A and $>$ 20 A) can follow shock impacts, which can pose serious risks to ground artificial conductors in short-, mid-, and long-term regimes \citep{Allen1989,Beland2005,Gaunt2007,Oliveira2018b}. Therefore, our results are consistent with previous reports showing nearly symmetric compressions generally cause more intense ground \dbdt{} variations associated with GICs \citep{Oliveira2018b,Xu2020a,Oliveira2021b,Oliveira2024a}. \par 

	Our results clearly show that GICs are enhanced promptly in the following 20 minutes after shock impacts, particularly resulting from nearly frontal shock impacts. This is consistent with the works of \cite{Oliveira2018b} and \cite{Xu2020a}, but for ground \dbdt{} variations. We then show for the first time that most intense GIC peaks ($>$ 5 A) occurred in the noon/dusk region (Figure \ref{mantsala_hist}) mostly due to nearly frontal shocks, with a very few being caused by highly inclined shocks. This noon/dusk preference is consistent with \cite{Madelaire2022a}, who showed that ground magnetic field responses to solar wind dynamic pressure enhancements usually occur in the dusk sector $\sim30$ minutes after the pressure pulse onset. The authors attributed this effect to partial ring current \citep{Fukushima1973} enhancements at high latitudes including \man's latitudes. In another work, \cite{Madelaire2022b} used ground magnetometer data and a model to derive equivalent ionospheric currents in response to solar wind dynamic pressure enhancements. The authors showed that a current vortex was mostly localized on the high-latitude duskside whereas a current vortex on high-latitude dawnside travelled with a significant velocity westward. These results also agree with the localized GIC peaks located on the duskside resulting from shock compression effects, being amplified by nearly frontal shock impacts, as shown by our case comparisons (Figures \ref{mantsala_598} and \ref{mantsala_307}) and statistical results (Figure \ref{mantsala_hist}). This is also shown in higher intensifications of the SuperMAG regional SMU index in the same region (MLT $\sim$ 18 hours) in response to impact of all shocks regardless of inclinations, but the SMR response becomes more enhanced as the shock becomes more frontal. These results will be reported in a forthcoming work.  \par

	\cite{Ngwira2018c} and \cite{Oliveira2021b} showed that very intense ground \dbdt{} variations are highly correlated and coincident in time with energetic particle injections originated in the magnetotail. The authors used spacecraft observations (Time History of Events and Macroscale Interactions during Substorms and Los Alamos National laboratory) located on the nightside tail around local magnetic midnight which were magnetically conjugated with ground magnetometers in North America. All stations and satellites were located a few hours around MLT = 00hr. \cite{Ngwira2018c} and \cite{Oliveira2021b} observed intense \dbdt{} variations occurring a few seconds after sharp and intense energetic particle injections observed by the spacecraft. The authors also noted intense aurora brightening associated with substorm occurrence and its subsequent poleward expansion of the auroral oval. These works associated these energetic particle injections to the tailward stretching of the local geomagnetic field at the magnetic midnight, usually caused by substorm-time flux growth phase dropouts \citep{Sauvaud1992,Reeves2001}. Additionally, \cite{Oliveira2021b} showed that the effects described above were more intense and occurred earlier in the case of a NFS in comparison to a HIS. These works support our findings concerning GIC peaks during substorm times occurring near the local magnetic midnight associated with substorm effects. This is clearly shown in our case examples (Figures \ref{mantsala_547} and \ref{mantsala_392}) and superposed epoch analysis (Figure \ref{mantsala_hist}D and Figure \ref{manpie}D).

	\add{\protect{As kindly suggested by a reviewer, we looked at time intervals between shock onsets and GIC peaks for all events ($\Delta$MLT = MLT2 -- MLT1 in Tables \ref{table1} and \ref{table2} for shock and substorm effects) as a function of \thxn. As shown in the L panels of Figures \ref{mantsala_598} and \ref{mantsala_307}, there is a time delay of $\sim$3 min and $\sim$7.5 min in MLT between shock onsets and GIC peaks for the NFS1 and HIS1, respectively. This is supported by a comparative study \citep{Oliveira2021b} and statistical analyses \citep{Oliveira2024a} showing that \dbdt{} variations are more intense and occur earlier as a result of nearly head-on shock impacts in comparison to highly inclined shock impacts. However, a clear correlations between $\Delta$MLT and \thxn{} were not found in our study. As mentioned before in this article, many works focusing on observations and simulations found a significant correlation between sudden impulse rise times and shock impact angles, with the shorter the rise time, the more frontal the shock impact \citep{Takeuchi2002b,Guo2005,Wang2006a,Selvakumaran2017,Rudd2019}. Therefore, those works report on magnetopause current effects recorded by ground magnetometer located at low- and mid-latitude regions. However, the GIC effects reported in this work were most likely caused by auroral electrojet dynamics above the \man{} compression station location \citep{Wawrzaszek2023}. Thus, the current data set we have in hands does not show strong correlations between GIC peak time delays and shock impact angles preasumably due to two reasons: 1) data collected at only one location is not enough to successfully address such correlation effetcs; and 2) such strong correlations do not occur at all. Therefore, more GIC measurements recorded at mid and low latitudes (e.g., New Zealand, Brazil, United States, Europe) are needed to address such possible correlation effects.}}

	Furthermore, we strongly recommend modelers to simulate the impact of IP shocks with different inclinations on the Earth's magnetosphere in GIC-related investigations. For example, \cite{Welling2021} simulated the impact of the ``perfect" coronal mass ejection on the Earth's magnetosphere suggested by \cite{Tsurutani2014a} to investigate the subsequent ground \dbdt{} response. The authors noted that the \dbdt{} response was amplified by the purely head-on nature of the CME impact, which is a very particular case. However, for more realistic results, we encourage modelers to undertake simulations of IP shocks impacts with different orientations on the magnetosphere. Therefore, the combination of asymmetric ground \dbdt{} variations \citep{Oliveira2018b,Oliveira2021b,Oliveira2024a} and varying ground conductivity \citep{Viljanen2017,Liu2019a,Wang2021} will most likely produced more realistic results, since most shocks detected in the solar wind at 1 AU have moderate inclinations of nearly $\sim$130$^\circ$-140$^\circ$ with respect tot he Sun-Earth line \citep{Oliveira2023c,Oliveira2024a}.

\section{Summary and conclusion}

	In this work, we used a subset with 332 events of a larger data set (603 events) of IP shocks from January 1999 to May 2023. We looked at IP shock impact angle effects on GICs recorded at the \man{} natural gas pipeline in southern Finland during two distinct moments: up to 20 minutes after shock impacts, due to shock compressions, and between 20 and 120 minutes after shock impacts, due to magnetotail activity. We summarize our findings as follows:

		\begin{enumerate}
			\item IP shock impact angles control GIC response at \man{}: nearly frontal shocks tend to trigger more intense GIC peaks. This is supported by previous observations of ground \dbdt{} response to shocks with different orientations \citep{Oliveira2018b,Xu2020a,Oliveira2021b,Oliveira2024a}.

			\item GIC peaks greater than 5 A tend to occur more after shock impacts. These peaks occur more around dusk as a response to nearly frontal shock impacts. These effects are explained by the enhancement of the partial ring current at \man{} latitudes in the dusk sector \citep{Madelaire2022a} associated with localized current vortices located around the dusk sector \citep{Madelaire2022b}.

			\item Very intense GIC peaks ($>$ 20 A) occur during substorm times around the magnetic midnight terminator. This occurs due to geomagnetic field line stretching around MLT = 00 hr in geospace caused by flux growth phase dropouts \citep{Sauvaud1992,Reeves2001}. Such flux dropouts generate intense energetic particle injections which in turn cause intense auroral brightening and intense ground \dbdt{} variations nearly simultaneously \citep{Ngwira2018c,Oliveira2021b}. Our GIC peak observations are also supported by intense \dbdt{} variations caused by nearly frontal shocks around the magnetic midnight \citep{Oliveira2021b,Oliveira2024a}.
		\end{enumerate}

\section*{Data Availability statement}
	
	Solar wind parameter and IMF data were downloaded from the Coordinated Data Analysis Web website (\url{https://cdaweb.gsfc.nasa.gov}), and processed according with the methodology described in \cite{Oliveira2023c}. The IP shock data base is available at the Zenodo repository \url{https://zenodo.org/records/7991430}. The SuperMAG data is located at \url{https://supermag.jhuapl.edu}. The GIC data recordings from the Finnish natural gas pipeline are located at \url{https://space.fmi.fi/gic/index.php?page=home}. Ground magnetometer data (Nurmij\"{a}rvi station) was downloaded from the IMAGE project website: \url{https://space.fmi.fi/image/www/index.php?page=home}. Daily sunspot number data is provided by the Sunspot Index and Long-term Solar Observations (SILSO) of the Royal Observatory of Belgium (\url{https://www.sidc.be/SILSO/datafiles}).

\section*{Conflict of Interest Statement}

	The authors declare that the research was conducted in the absence of any commercial or financial relationships that could be construed as a potential conflict of interest.

\section*{Author Contributions}

	This research article was written by the author based on discussions with the co-authors. The co-authors read, commented, and approved this manuscript.

\section*{Funding}
	
	DMO and EZ thank the financial support provided by the NASA HGIO program through grant 80NSSC22K0756. DMO and SVL acknowledge the financial support provided by NASA LWS program through grant NNH22ZDA001N-LWS.

\bibliographystyle{frontiersinSCNS_ENG_HUMS} 
\bibliography{Oliveira_main}


\end{document}